\documentclass[fleqn,10pt]{wlscirep}
\usepackage{amsmath}

\usepackage{bm}
\title{Unsharpness of generalized measurement and its effects in entropic uncertainty relations}

\author[1]{Kyunghyun Baek}
\author[1,2*]{Wonmin Son}
\affil[1]{Department of Physics, Sogang University, Mapo-gu, Shinsu-dong, Seoul 121-742, Korea}
\affil[2]{University of Oxford, Department of Physics, Parks Road, Oxford OX1 3PU, United Kingdom}

\affil[*]{sonwm71@sogang.ac.kr}

\begin{abstract}
Under the scenario of generalized measurements, it can be questioned how much of quantum uncertainty can be attributed to measuring device, independent of the uncertainty in the measured system. On the course to answer the question, we suggest a new class of entropic uncertainty relation that differentiates quantum uncertainty from device imperfection due to the unsharpness of measurement. In order to quantify the unsharpness, we {suggest} and analyze the quantity that characterizes the uncertainty in the measuring device, based on Shannon entropy. Using the quantity, we obtain a new lower bound of entropic uncertainty with unsharpness and it has been shown that the relation can also be obtained under the scenario that sharp observables are affected by the white noise and amplitude damping.
\end{abstract}
\begin{document}

\flushbottom
\maketitle
\thispagestyle{empty}

\section*{Introduction}
After Heisenberg introduced the uncertainty relation \cite{Heisenberg1927}, it has become the central principle in quantum physics and it still gives rise to subsequent debates in the clarification of its underlying meaning todate \cite{Busch2007}. It is well-known that the early formulation of the uncertainty relations have been based upon the statistical quantification as like standard deviation. The uncertainty relation (UR) has been widely {expressed} in the form of Robertson's UR \cite{Robertson1929} as 
\begin{eqnarray}\centering\label{RUR}
\Delta A\Delta B\geq \frac{1}{2}|\langle\psi|[\hat{A},\hat{B}]|\psi\rangle| 
\end{eqnarray}
for two self-adjoint operators $\hat A$ and $\hat B$ which apply to a quantum state $|\psi\rangle$. The notations are $[\hat A,\hat B]=\hat A\hat B-\hat B\hat A$ and $(\Delta A)^2=\langle\hat{A}^2\rangle-\langle \hat A\rangle^2$. Standard deviation, however, is not the ultimate measure of uncertainty, in the sense that well-defined measure need to be invariant under the relabeling of measurement outcomes as discussed \cite{Uffink1990}. Moreover, the lower bound in \eqref{RUR} is not only state-dependent, but also vanishes for states $|\psi\rangle$ which are not common eigenstates of non-commuting operators \cite{Deutsch1983}.
{Overcoming the incompleteness}, Shannon entropy was to be used in order to formulate UR which generates so-called entropic UR. As a result, it is found that entropic UR is stronger than that of Robertson's UR \eqref{RUR} for the case of continuous variables systems\cite{Biaynicki1975}. On the other hand, for discrete variables, the entropic UR comes to have a state-independent bound\cite{Uffink1988} and it is shown that the entropic UR is neither stronger nor weaker than (\ref{RUR}) in general \cite{Baek2014}.

Here, we would like to point out that all the previous discussions are under the assumption of projection-valued measure (PVM) in their measurements. {They do not} encompass the general circumstance of quantum measurements. The generalized measurements are represented by positive-operator-valued measures (POVMs) which are set of positive operators, satisfying the completeness but not orthogonality necessarily. {To specify the orthogonality, we shall call an observable {\it sharp} if it can be described by PVM, otherwise we call it {\it unsharp}. Concept of unsharp observables plays important roles in quantum information as they can be used to extract more information from a quantum system than measurements described by PVM \cite{Peres1991}. }
 
The entropic UR for POVMs has been formulated by Krishna {\it et al}\cite{Krishna2002} for the first time. Later, alternative form of the formula has been proposed \cite{Tomamichel2012} and it is proved that the lower bound is to be stronger than the one in Ref. \citenum{Krishna2002}, as following \cite{Coles2014}. Consider two observables $A$ and $B$ described by POVMs $\{\hat A_i\}$ and $\{\hat B_j\}$, respectively. Then, Shannon entropy is defined as $H_\rho(A)=-\sum_i p^A_i \log_2 p^A_i$ associated with probability distributions $\{p_1^A,p_2^A,...,p_m^A\}$ where $p^A_i=\text{Tr}[\hat\rho\hat A_i]$ (similarly for $H_\rho(B)$). As a matter of brevity, base of $\log$ will be omitted through this work. Using the operator norm $\|\hat A\|:=\max\{\|\hat A|\xi\rangle\| : \| |\xi\rangle\|=1\}$, i.e. maximal singular value of $\hat A$, the entropic UR is of the form\cite{Tomamichel2012,Coles2014}
\begin{align}\label{EUR}
H_{\rho}(A)+H_{\rho}(B)\geq -\log C
\end{align}
for a quantum state $\hat\rho$, where the lower bound is defined as 
\begin{align}
C=\min\left[\max_i\| \sum_j \hat B_j\hat{A_i}\hat B_j\|,\max_j\|\sum_i \hat A_i \hat B_j \hat A_i\|\right].
\end{align}
From its definition, the bound $C$ is given independent to the state and the relation (\ref{EUR}) is reduced to the Massen and Uffink UR \cite{Uffink1988} for projective measurements. For the properties, one can refer the work in Ref. \citenum{Coles2015} for further details. In the similar vein, the entropic uncertainty relations have been studied with use of generalized entropies in \citenum{Rastegin2011}. The effects of entanglement and mixedness on the lower bounds of entropic UR have been discovered\cite{Berta2010} and discussed extensively \cite{Coles2011, Coles2014}. Although generalized entropic UR for POVMs has been derived and discussed before, the general operational meaning of the bound is not very clear in the literatures. When the measurements are projective, uncertainty relations impose the trade-off constraint of two ideal measurements. In comparison, when the measurement is POVM, the relationship includes the additional uncertainties coming from the unsharpness of measuring devices in its lower bound. In the inequality, thus, uncertainties either from quantum system or unsharpness of measuring devices should be identified in principle. Here, we provide the quantification that the different source of "uncertainties" can be discriminated in general.

In the preliminary works, quantifications of intrinsic unsharpness of measuring device described by POVM have been formulated using operator norm and the role of unsharpness on joint measurability has been investigated \cite{Heinosaari2008, Busch2009}. Especially in the work by Busch\cite{Busch2009}, axioms that a measure of unsharpness should obey has been proposed. From a different angle, Massar has also defined additional uncertainty coming from the intrinsic unsharpness in terms of statistical variance\cite{Massar2007}. In the same vein, here, we try to quantify the measure of unsharpness in a measuring device based on entropy. Our quantification averages the entropic uncertainties in the measuring process and is called {\it device uncertainty}. From the definition, one can naturally discriminate the device uncertainty from the uncertainty in the original state and we attribute the latter to {\it quantum uncertainty}. Furthermore, we provide a way to quantify the device uncertainty under POVM and differentiate it from the quantum uncertainty in the measured system. With the quantifications, we investigate the explicit effects of unsharpness of POVM on entropic URs.

This paper is organized as following. Device uncertainty is defined as a measure of unsharpness based on entropy, and its appropriate properties is provided. Physical meaning of the device uncertainty, then, is clarified as we analyze two-level systems. Using the definition of device uncertainty, we introduce quantum uncertainty and show that it satisfies relevant properties. Based on quantifications of unsharpness, effects of unsharpness on entropic uncertainty is considered for specific noise models such as white noise and amplitude damping.

\subsection*{Known lower bound of entropy for a single observable}
Before we quantify unavoidable uncertainty originating from unsharpness of measuring device, let us consider the entropic lower bound  for single observable $A$. It is described by set of operators $\{\hat A_i\}$ which are positive definite satisfying $\sum_{i=1}^n\hat A_i=\hat I_d$. Here, $n$ is the number of outcomes and $\hat I_d$ denotes the identity matrix for $d$-dimensional system. With the POVM, one cannot predict measurement outcomes deterministically over all the prepared states in general. The situation can be described by entropic bound as follows 
\begin{align}\label{single}
H_{\rho}(A)\geq -\log\left[ \max_i\|{\hat A_i}\|\right],
\end{align}
which is originally provided { by Krishna {\it el al}}\cite{Krishna2002}. The lower bound for single observable reflects the imperfection of the measuring device $A$. It means that, when the outcomes are obtained from a measuring device with imperfect resolution, uncertainty of the original system in terms of entropy is lower-bounded by the most accurate resolution scale among the outcomes. 

\begin{figure}[t]
\begin{minipage}[h]{1.0\linewidth}\centering
\includegraphics[scale =0.5]{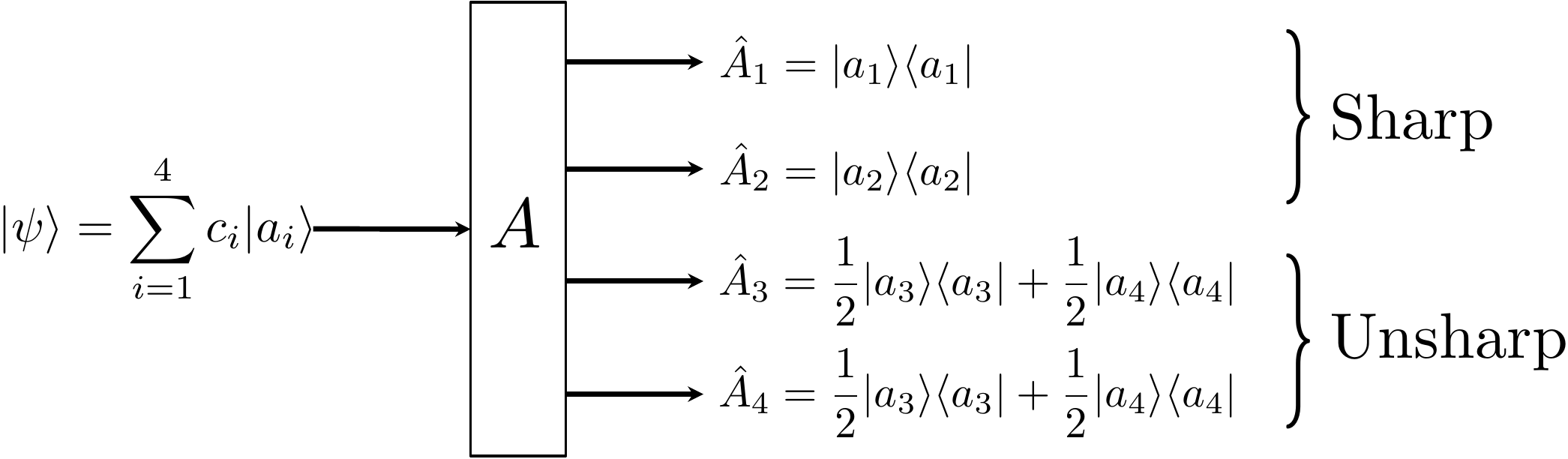}
\end{minipage}
\caption{A schematic exemple for the experiment of generalized measurement $A$ described by $\{\hat A_i\}$ acting on 4-dimensional quantum system. This figure illustrates a nontrivial case of unsharp measurement $A$ that is partially sharp, where input state $|\psi\rangle$ is expressed as the linear superposition of orthonormal states $\{|a_i\rangle\}$ with complex numbers $c_i$ obeying $\sum_{i=1}^4 |c_i|^2=1$. The input state, then, can be sharply measured if $c_3=c_4=0$. Otherwise, we can not sharply measure it, since we can not distinguish $|a_3\rangle$ and $|a_4\rangle$.}
\label{Unsharpness}
\end{figure}

The lower bound of (\ref{single}) can vanish when one of the elements of POVM is a projector even if the other POVM elements are not full projective. The situation is depicted in Figure \ref{Unsharpness}. It is when the partial sectors of a measuring device are projective while the rest are fuzzy as it is delineated in Figure \ref{Unsharpness}. Thus, the lower bound of entropy can disappear even though the measuring device is not perfect. It implies that the lower bound does not appropriately quantify the extent of device uncertainty coming from measurement unsharpness. In order to quantify and distinguish device uncertainty in terms of entropy, we need to identify more proper quantification of genuine lower bound of entropy in the measuring device. Here, we suggest a new measure of device imperfection and examine its properties in the following.

\section*{Results}

\subsection*{Device uncertainty for general POVM}
Let us derive a stronger lower bound of entropy than the one in (\ref{single}). We show that it is well-formulated as a measure of unsharpness at the same time. 
As a first step, let us assume that a POVM of an observable $A$ is described in $\mathcal{H}_d=\mathbb{C}^d$ of which the number of outcomes is $n$. Each element of $A$, then, can be written as
\begin{align}\label{POVM2}\centering
\hat A_i=\sum_{k=1}^d a^k_{i} |a^k_{i}\rangle\langle a^k_{i}|,
\end{align}
where $a^k_{i}$'s are eigenvalues of $i$-th element of POVM. $|a^k_{i}\rangle$ is an eigenstate corresponding to $a^k_{i}$ satisfying $0\leq a^k_{i}\leq1$ due to the positivity and completeness relation. 
{ Unsharpness of $\hat A_i$ disappears only when they are written as projectors.  By means of the expression \eqref{POVM2}, the projective condition is equivalent to the statement that all eigenvalues of $\hat A_i$ are unity as they are given by $a^k_{i}\in \{0,1\}$ \cite{Busch2009}. From the fact, it can be argued that $-\log\max_i \|{\hat A_i}\|$ solely cannot be an appropriate measure of unsharpness. It is because $-\log\max_i\|{\hat A_i}\|=0$ does not mean that $\hat A_i$ for $\forall i$ are projectors. To overcome this, one can identify that an unsharpness measure should be defined as a function of eigenvalues $a_i^k$ vanishing when $a_i^k$ are given by 0 or 1. Additionally, in order to capture nontrivial cases that an unsharp measurement is partially sharp as depicted in Figure \ref{Unsharpness}, it is necessary to take into account how much overlap exists between a given state $\hat\rho$ and $|a_i^k\rangle$. 
 As a specific function reflecting these features, let us take $h(a_i^k)=-a_i^k\log a_i^k$ and take the average such that $\sum_{k=1}^d \langle a_i^k|\hat\rho|a_i^k\rangle h(a_i^k)$. Finally, summing over all the POVM elements, we obtain 
 \begin{align}\label{DU}
 D_{ \rho}(A)=\sum_{i=1}^n\left(\sum_{k=1}^d \langle a^k_i|\hat\rho|a^k_i\rangle h(a_i^k)\right),
\end{align}
which is called as device uncertainty. It can be proved that the entropy of probability distribution $A$ is lower bounded by the quantity, $D_\rho(A)$, due to the concavity of the log function $h(x)$\cite{Cover1991} applied to $p^A_i=\sum_{k=1}^d a_i^k \langle a^k_i|\hat\rho|a^k_i\rangle$.
Furthermore, the device uncertainty can be proved to be a stronger lower bound than the one in (\ref{single}). Namely, their relationship is written as
\begin{align}\label{single2}
H_\rho(A)\geq D_\rho(A) \geq \min_{\rho} [D_\rho(A)] \geq -\log \max_i\|{\hat A_i}\|,
\end{align}
where the minimal device uncertainty over states is obtained by diagonalizing $\sum_{i=1}^n\sum_{k=1}^d h(a_i^k)|a_i^k\rangle\langle a_i^k|$ and taking the lowest eigenvalue of it. Thus, the device uncertainty defined as an unsharpness measure gives us not only lower bound of the entropy by itself, but also improved state-independent lower bound.
Detailed proof of third inequality can be found in Method section.

Now, let us show that the device uncertainty is appropriate measure for the unsharpness of POVM. 
In the previous work, Massar has defined a quantity to characterize an additional uncertainty coming from intrinsic unsharpness of POVM based on statistical variance\cite{Massar2007}. Subsequently, he has proposed a list of criteria to show the quantity is appropriately defined as a measure of unsharpness. In accordance with the approach, we prove the validity of the device uncertainty as follows.}

\begin{itemize}
\item[($D$-i)] $H_\rho(A)\geq D_\rho(A,)\geq 0$, which can be found due to the concavity of entropy \cite{Cover1991}.

\item[($D$-ii)] $D_\rho(A)=H_\rho(A)$ for all states if and only if $\hat A_i=\lambda_i \hat I_d$ with $0\leq\lambda\leq1$ satisfying $\sum_i \lambda_i=1$. Namely, the entropic uncertainty of measurement outcomes distribution only comes from the device uncertainty, $D_\rho(A)$.

\item[($D$-iii)] $D_\rho(A)=0$ for all states if and only if $A$ is PVM.

\item[($D$-iv)] A convex combination of two POVMs cannot be sharper than these two POVMs themselves. For example, let us consider two observables $A$, $B$ acting on same quantum system described by $|\psi\rangle$ with probability $p$ and $q$ satisfying $p+q=1$, respectively. We construct the convex combination of two POVMs $\{p\hat A_i, q\hat B_j\}$, then the device uncertainty becomes larger than the convex combination of device uncertainty such as 
\begin{align*}
D_\rho\big(pA+qB\big)=pD_\rho(A)+qD_\rho(B)+H_{\text{bin}}(p),
\end{align*}
where the binary entropy is defined by $H_{\text{bin}}:=-p\log p-q\log q$
\end{itemize}

{ We have shown that $D_\rho(A)$ satisfies the criteria for unsharpness. there are a couple of comments that can be made on the criteria. Firstly, the upper bound of the device uncertainty is given by $H_\rho(A)$ and the gap quantifies how much entropic uncertainty originates from the state. Secondly, an important point of the quantity is that the device uncertainty is state-dependent. In our case, we consider the additional uncertainty from unsharpness that is state-dependent as depicted in Figure \ref{Unsharpness}. It is the case when the device imperfection is differently responding to the measured state as the degree of unsharpness can be varied depending upon the state.
For instance, when $c_1=c_2=1/\sqrt{2}$ for the state $|\psi\rangle=\sum_{i=1}^{4}c_i |a_i\rangle$ in Figure \ref{Unsharpness}, entropy of the outcome probabilities is given as $\log2$ as the state is corresponding to the sector of sharp measurements. In the case, due to the projective property of the corresponding sector, we have $D_\rho(A)=0$. On the other hand, when $c_1=c_2=0$, $c_3=c_4=1/\sqrt{2} $, entropy is also given as $\log2$ and the uncertainty originate from unsharpness of measurement as to give $D_\rho(A)=\log2$. As a way to quantify state-independent unsharpness in terms of device uncertainty, the maximally mixed state can be considered as a measured state, covering all degrees of freedom. }

\begin{figure}[t]\centering
\includegraphics[scale =0.6]{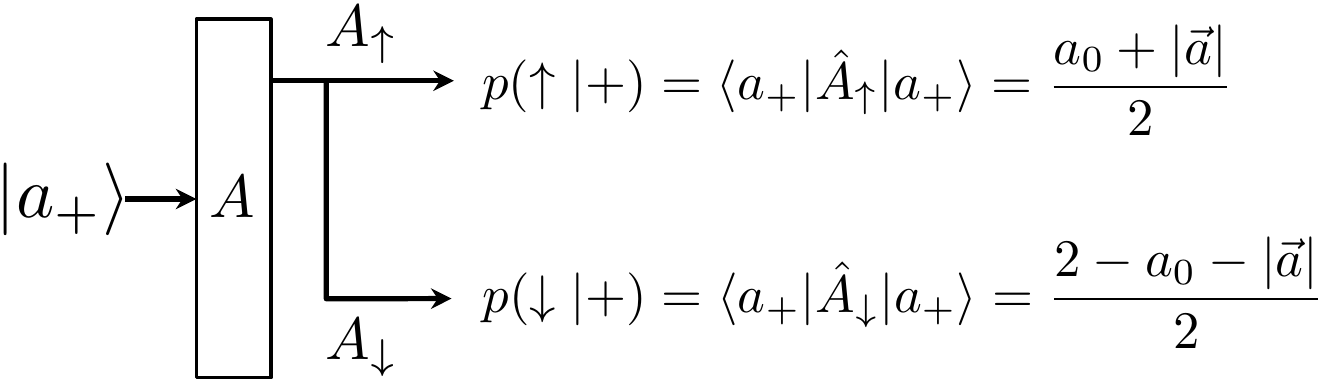}
\caption{Scheme illustrates a situation that the eigenstate corresponding to positive value, $|a_+\rangle$, is injected into the measurement $A$. In this case, the probabilities to obtain $\uparrow$ and $\downarrow$ are denoted by the conditional probability $p(\uparrow|+)$ or $p(\downarrow|+)$. Then, in this case, the probability for bit flip error corresponds to $p(\downarrow|+)$ and the uncertainty emerged by the bit flip error is quantified by $H_{bin}\Big(p(\uparrow|+)\Big)$. In the same way, a situation that $|a_-\rangle$ is injected can be considered.}\label{Scheme}
\end{figure}

\subsection*{Device uncertainty for two-level system}
In order to clarify the meaning of the device uncertainty, let us consider a POVM $\hat{A}$ in two-level quantum system which is described in two dimensional Hilbert space, $\mathcal{H}_2$. We define their elements as
\begin{equation}\label{qubit}
\hat A_\uparrow=\frac{1}{2}\left(a_0\hat I_2 +  \; \vec{a}\cdotp\vec{\sigma}\right)~ \mbox{and}~ \hat A_\downarrow=\hat I_2- \hat A_\uparrow,
\end{equation}
where $|\vec a|\leq a_0\leq 2-|\vec a|$. In this case, the operators, $\hat A_{\uparrow,\downarrow}$, are decomposed into bases $|a_\pm\rangle$ which are the eigenstates of $\vec{a}\cdot \vec{\sigma}$ and their corresponding eigenvalues are denoted by 
\begin{align*}
\langle a_\pm|\hat A_{\uparrow,\downarrow}|a_\pm\rangle=p(\uparrow,\downarrow|\pm),
\end{align*}
where $p(\uparrow,\downarrow|\pm)$ are the conditional probabilities of the measurement outcomes, $\uparrow,\downarrow$, for the input states $|a_\pm\rangle$, respectively. The notations are sketched in Figure \ref{Scheme}. For the case of an ideal measurement, i.e. $a_0=|\vec{a}|=1$, which results $p(\uparrow|+)=1$ and $p(\downarrow|+)=0$. In general, the conditional probabilities take arbitrary values between $0$ and $1$. It follows that, among the conditional probabilities and the parameters, there are relationships such as $p(\uparrow|\pm)=(a_0\pm|\vec{a}|)/2=1-p(\downarrow|\pm)$. Then, it is trivial to identify
\begin{align*}
H_{\text{bin}}\Big(p(\uparrow|+)\Big)=H_{\text{bin}}\Big(p(\downarrow|+)\Big)
\end{align*}
where $H_{\text{bin}}$ is binary entropy defined as $H_{\text{bin}}(p)=-p\log p-(1-p)\log (1-p)$. Using the relation, according to (\ref{DU}), we can evaluate the device uncertainty of $A$ for a input state $|\psi\rangle$ as follows,
\begin{align}\label{DUqubit}
D_\psi(A)=\sum_{i=\pm}|\langle \psi|a_i\rangle|^2 H_{\text{bin}}\Big(p(\uparrow|i)\Big).
\end{align}
Only when the measurement is sharp, this value vanishes for all states, so thus there is no uncertainty from the imperfection of measuring device. Physically, the device uncertainty can be interpreted as the averaged  quantification of the {\it bit flip error} in the measuring device. It is because the probability for the bit flip error is given by $p(\downarrow|+)$ implying that the measuring device misjudge the state when the input state is the eigenstate $|a_+\rangle$. It can be evaluated that $H_{\text{bin}}\Big(p(\downarrow|+)\Big)$ is maximized when $p(\downarrow|+)=1/2$.
In the same way, we can obtain the device uncertainty for the orthogonal state $|a_-\rangle$ whose value can be different from the state $|a_+\rangle$ in general. Thus, the device uncertainty in (\ref{DUqubit}) is interpreted as the averaged value of the device imperfection with respect to arbitrary state $|\psi\rangle$.
Even though it is averaged quantity, $D_\psi(A)$ is sensitive to the measured state. It is because the measuring device can be responded differently for different input states in the most general scenario. 

\subsection*{Differentiated entropy for the uncertainty in measured state.}
\begin{figure}[t]
\centering
\includegraphics[scale =0.35]{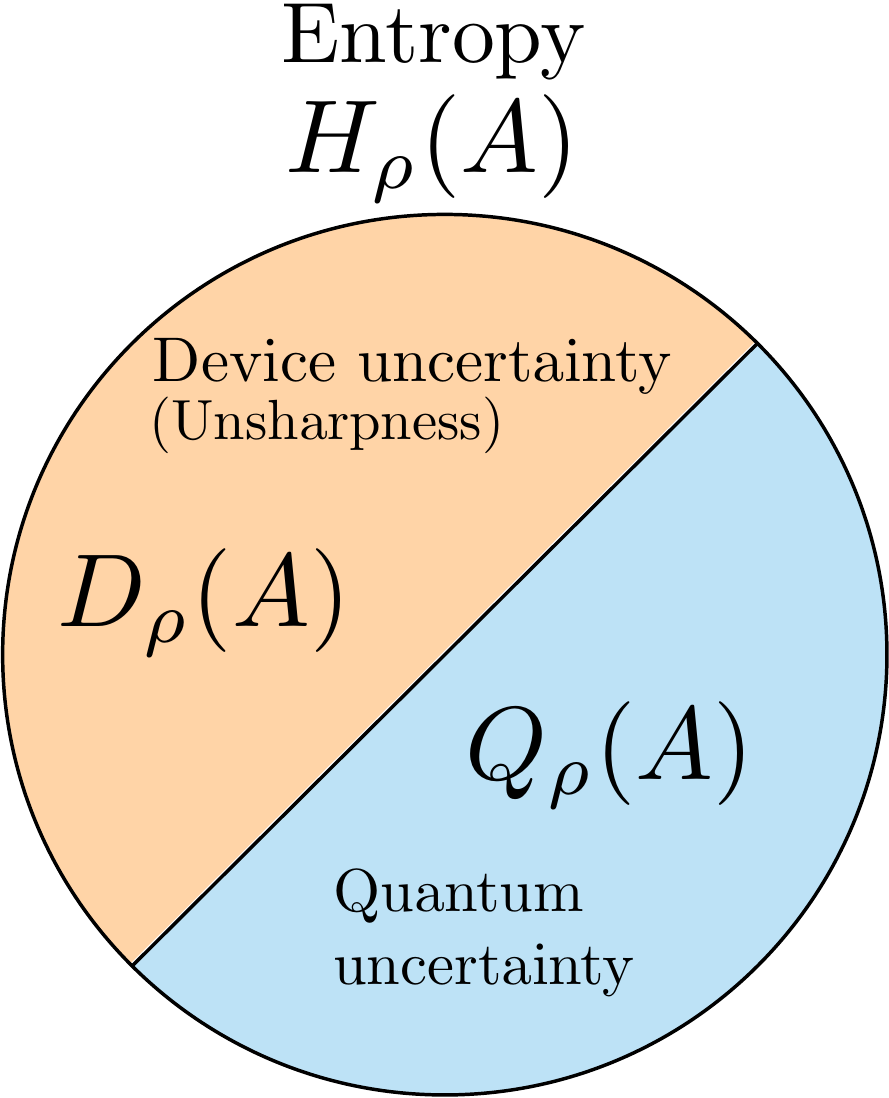}
\caption{Diagrams illustrates the relation among entropy, device and quantum uncertainty. Sum of the device and the quantum uncertainties are the same with entropy. Thus, we distinguish the uncertainty according to where it comes from.}\label{DQ}
\end{figure}

Once we have full characterization of the uncertainty in the measuring devices, it is possible to extrapolate the amount of uncertainty in the original quantum system. Due to the additive nature of entropy, the uncertainty in the original system can be characterized by subtraction of device uncertainty from entropy of the final outcomes, as follows
\begin{align}\label{SU}
Q_\rho\left(A\right)=H_\rho\left(A\right)-D_\rho\left(A\right).
\end{align}
Related with the properties, one can find that the quantum uncertainty is equal to entropy when device uncertainty becomes trivial, i.e. $D=0$. As the unsharpness of the measuring devices increases, the information that we can extract from the original system decreases so that $Q$ becomes smaller. Thus, the quantity $Q$ characterizes the amount of uncertainty in the measured system whose properties can be summarized as follows.
\newline
\begin{itemize}
\item[($Q$-i)] $0\leq Q_\rho(A)\leq H_\rho(A)$, which can be found due to the condition (D-i).
\item[($Q$-ii)] $Q_\rho(A)=H_\rho(A)$ for all states if and only if $A$ is PVM due to the condition (D-ii).
\item[($Q$-iii)] $Q_\rho(A)=0$ for all states if and only if $\hat A_i=\lambda_i \hat I_d$ with $0\leq\lambda\leq1$ satisfying $\sum_i \lambda=1$ due to the condition (D-iii).
\item[($Q$-iv)] A convex combination of two POVMs cannot increase the quantum uncertainty. This is also trivial by using the condition (D-iv). An increase of entropy is same with an increase of the device uncertainty by $H_{\text{bin}}(p)$ for  the convex combination of two POVMs. Thus, as they are canceled each other, the quantum uncertainty is invariant.
\end{itemize}
Thus, using the quantities, we can divide the total uncertainty characterized by entropy into device and quantum uncertainties as depicted in Fig. \ref{DQ}. 

\subsection*{Device uncertainty and entropic UR under white noise}
Entropic UR \eqref{EUR} tells us that there is a fundamental limit to prepare a state providing definitive outcomes of two incompatible observables simultaneously. Under the consideration of POVM, however, unsharpness of measuring devices affects lower bounds of the entropic UR as the device uncertainty becomes the lower bound of entropy for single observable. Due to the additive nature of entropy, summation of device uncertainties for more than two observable can also be taken as a lower bound of the composite entropies. In other words, unsharpness affects on lower bounds of entropic UR in an additive way. In order to investigate the effects of unsharpness on lower bound of entropic URs, we will consider an observable with an addition white noise, as follows.

Let us consider unsharp measurements $A_{\alpha}$ constructed by adding white noise on sharp measurements in $d$-dimensional quantum system. Thus each positive operator corresponding to $i$-th outcome of measurement $A$ is defined in the form of
\begin{align}\label{WN}
\hat A_{i}(\alpha)=\alpha|a_i\rangle\langle a_i|+(1-\alpha)\frac{\hat I_d}{d}
\end{align}
with a mixedness parameter $0\leq\alpha\leq1$. The white noise has been taken into account as a representative noise model, especially when we discuss joint measurability\cite{Heinosaari2015,Uola2014,Karthik2015}. In this case, device uncertainty has a value as follows,
\begin{align}\label{DC}
{D}(A_{\alpha})=-(\alpha+\alpha_d)\log(\alpha+\alpha_d)-(d-1)\alpha_d\log\alpha_d
\end{align}
where $\alpha_d\equiv(1-\alpha)/d$. The white noise acts equally on measurements regardless of states, so that the device uncertainty is only determined by $\alpha$, i.e. state-independent. It can be found that the device uncertainty is monotonically decreasing function of $\alpha$ and the behavior is homogenous for different dimensions $d$. In the following, we will derive lower bounds of entropic URs for these unsharp measurements.

Let us consider two unsharp measurements $A_{\alpha}$ and $B_{\beta}$ constructed under the model of white noise that acts on arbitrary orthonormal bases  $\{|a_i\rangle\}$ and $\{|b_j\rangle\}$ as in (\ref{WN}) respectively. Relation between them is characterized by the inner product of the bases $U_{ij}=\langle a_i|b_j \rangle$. In that case, the composite entropies $H_{\rho}(A)+H_{\rho}(B)$ is bounded by the constraint of quantum incompatibility in addition to the device uncertainties. The lower bound can be obtained by straightforward induction
\begin{align} 
 &\mathcal B_{1}=\log c_{ab} +\min [D(A_{\alpha}),D(B_{\beta})], \label{Qwn}
\end{align}
with Massen-Uffink(MU) bound\cite{Uffink1988} $-\log c_{ab} =-\log \max_{i,j} |U_{ij}|^2$ determined by the inner product between the maximal orthonormal bases $\{|a_i\rangle\}$ and $\{|b_j \rangle\}$, of which proof is provided at the Method section.

The bound $\mathcal B_{1}$ in (\ref{Qwn}) is given by the addition of the smaller device uncertainty to MU bound. The bound $\mathcal B_1$, therefore, is written by the decomposition of two terms. The first term stands for incompatibility of two observables and the second term is to represent the device uncertainty from unsharpness. Additionally, the bound $\mathcal B_1$ is optimized for mutually unbiased bases as it is saturated by eigenstates of an observable to have bigger unsharpness. However, $\mathcal B_{1}$ does not take the optimal value at the other extreme. The bound does not saturate the lower bound in the limit of two identity POVMs, $\alpha=\beta=0$, having the identical bases $|a_i\rangle=|b_i\rangle \forall i$. It is because the bound have smaller values than the sum of  entropies, $H(A_{\alpha})+H(B_{\beta})$ whose lower bound is given by device uncertainties,
\begin{align} 
&\mathcal D_{WN}=D(A_{\alpha})+D(B_{\beta})
\end{align}
as to be consistent with (\ref{single2}). The bound $\mathcal B_1$ is, therefore, not well-saturated since device uncertainties give us a stronger bound itself even for incompatible orthonormal bases. In order to inspect the differences in detail, we consider more general bound of entropic UR using majorization relation in the following.

\subsubsection*{Entropic uncertainty relation under majorization relation}
We now show how to obtain an improved bound of entropic UR which is derived from the majorization relation of two incompaitible measurements. The bound is new characterization of entropic UR which factors out the contribution of device uncertainty and it provides the stronger quantification of entropic UR than the most recent one\cite{Rudnicki2014}. The majorization approach has been also extended to the cases with the generalized entropies recently \cite{Rastegin2015}  For the characterization, let us first introduce the recent entropic uncertainty based upon the majorization relation as
\begin{equation}\label{MajEUR}
H_{\rho}(A)+H_{\rho}(B)\ge H(W)
\end{equation}
{which is proposed for arbitrary orthonormal bases $A$ and $B$ described by $\{|a_i\rangle\}$ and $\{|b_j\rangle\}$, respectively. The lower bound $H(W)$ is Shonnon entropy of the majorized probability distribution $\vec W=(w_1-1,w_2-w_1,...,w_d-w_{d-1},0,...,0)$ whose elements are defined as }
\begin{align} 
w_k:= \max_{\substack{ \mathcal R, \mathcal S\\  |\mathcal R|+|\mathcal S|=k+1 }}\left\|{\sum_{i\in \mathcal R}|a_i\rangle\langle a_i|}+{\sum_{j\in \mathcal S} |b_j\rangle\langle b_j|}\right\|,
\end{align}
where $\mathcal R$ and $\mathcal S$ are subsets of $\{1,2,...,d\}$ and $|\mathcal R|$ denotes the number of components of $\mathcal R$ similarly for $|\mathcal S|$ (see Ref. \citenum{Friedland2013,Puchala2013} for detailed descriptions). Using these coefficients, the direct-sum majorization relation is written as 
\begin{align} \label{Majorization}
 \vec{p}^A\oplus \vec{p}^B\prec \{1\}\oplus \vec{W}
\end{align}
with the definition $\vec{p}^A\oplus \vec{p}^B$$=$ $(p^A_1,p^A_2,$$...p^A_d,p_1^B,p^B_2,$$...p^B_d)$,
where $\vec{r}\prec \vec{s}$ means that a $m$-dimensional vector $\vec{r} \in \mathbb R^m$ is majorized by a vector $\vec{s}\in\mathbb R^m$, i.e. $\sum_{i=1}^k r^\downarrow_i\leq \sum_{j=1}^{k}s^\downarrow_j$ for $1\leq k\leq m-1$ with $\sum_{i=1}^{m}r^\downarrow_i=\sum_{j=1}^{m}s^\downarrow_j$. The downarrow denotes the vector elements that are sorted in decreasing order, i.e. $r_1^\downarrow\geq r_2^\downarrow\geq...\geq r_m^\downarrow$.

Let us take into account the differentiated entropy of unsharp measurement $A_{\alpha}$ for the quantum uncertainty in (\ref{SU}). Under the consideration of the measurement with white noise in (\ref{WN}), the uncertainty in a quantum state can take the form, 
\begin{align} \label{Qwn}
 Q_\rho(A_{\alpha})=H_{\rho}(A_{\alpha})- D_{\rho}(A_{\alpha})=\sum_{i=1}^d f(p_i^A,\alpha)
\end{align}
where each probability is denoted by $p_i^A=\langle a_i|\hat \rho|a_i\rangle$, and the function $f$ is concave for $0\leq p\leq 1$ and $0\leq\alpha\leq1$. Explicit form of the function can be found as 
\begin{align} 
f(p,\alpha)=-(\alpha+\alpha_d)p \log\left[\frac{\alpha p+\alpha_d}{\alpha+\alpha_d}\right]-\alpha_d(1- p) \log\left[\frac{\alpha p+\alpha_d}{\alpha_d}\right].
\end{align}
For another measurement, saying $B_\beta$, the quantum uncertainty $Q_{\rho}(B_{\beta})$ can also be obtained in a similar manner. {In the case, it can be found that the quantum uncertainties $Q_\rho(A_{\alpha})$ and $Q_\rho(B_{\beta})$ are Schur-concave functions with respect to probability distributions $\vec p^A$ and $\vec p^B$, respectively. Subsequently, it is also possible to prove that they satisfy the following inequalities,
\begin{align} 
Q_\rho(A_{\alpha})+Q_\rho(B_{\beta})&=\sum_{i=1}^d f(p_i^A,\alpha)+\sum_{j=1}^d f(p_j^B,\beta)\\
&\geq \sum_{i=1}^d f(p_i^A,\min[\alpha,\beta])+\sum_{j=1}^d f(p_j^B,\min[\alpha,\beta])\\
&\geq \sum_{i=1}^{2d} f({W_i,\min[\alpha,\beta]})\equiv Q(W),
\end{align}
where the second line is obtained from a property $f(p,\alpha)\geq f(p,\beta)$ for $\alpha\geq\beta$ and the third line follows since Schur-concave functions preserve the partial order induced by majorization relation \eqref{Majorization} together with the fact that $f(1,\alpha)=f(0,\alpha)=0$. From the relations, an entropic URs is derived in the form of
\begin{equation}
H_{\rho}(A_\alpha)+H_{\rho}(B_\beta) \ge Q(W)+D(A_\alpha)+D(B_\beta)\equiv \mathcal B_2 \ge H(W)
\end{equation}
where $\mathcal B_2$ gives stronger characterization than $H(W)$ that becomes equal to $\mathcal B_2$ only when $\alpha=\beta=1$, i.e. both measurements are sharp. 
This bound depends on larger unsharpness, so that whenever at least one of measurements is extremely unsharp, i.e. $\min[\alpha,\beta]=0$, the bound $\mathcal B_2$ is reduced to $\mathcal D_{WN}$. 
Nevertheless, important point is that the bound $\mathcal B_2$ is always bigger than total device uncertainty $\mathcal D_{WN}$ for incompatible orthonormal bases, while $\mathcal B_1$ may not. }

\begin{figure}[t]\centering
\centering
\includegraphics[scale =0.33]{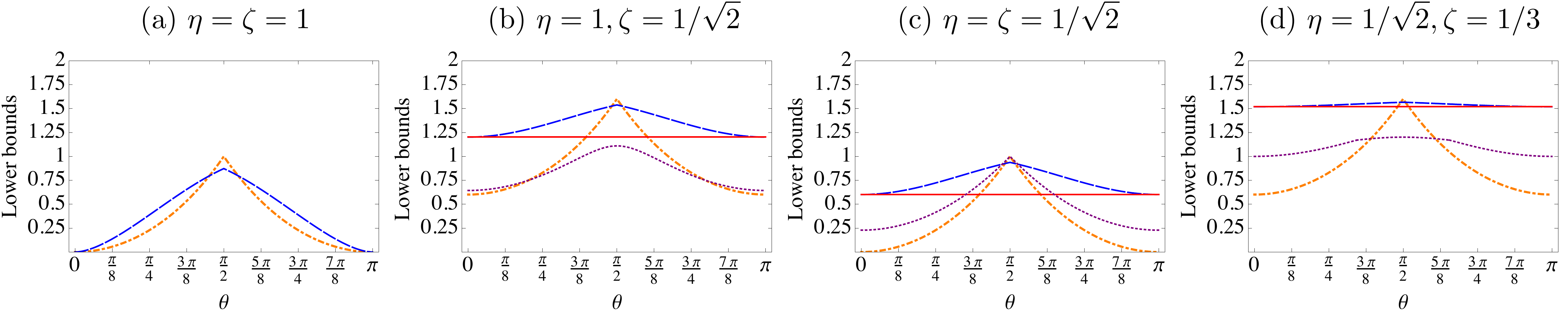}  
\caption{Graphs illustrate the lower bounds $\mathcal B_1$ (orange dot-dashed lines), $\mathcal B_2$ (blue dashed lines), $-\log C$ (purple dotted lines), and $\mathcal D_{WN}$ (red solid lines) for spin observables $X_\eta$ and $Z_\zeta$ with respect to angle $\theta$. In Figure (a), we denote $-\log C$ and $\mathcal B_1$ together by orange dot-dashed lines. And four the other graphs are sequentially presented for different unsharpnesses, and accordingly it is observed that the bounds become larger as unsharpness increases. We can analyze their behaviors as comparing the bounds to device uncertainty, and it is identified that $-\log C$ is strictly lower than device uncertainty at large unsharpness.} 
\label{QDwhite}
\end{figure}

{In order to clearly see the effect of unsharpness on lower bounds $\mathcal B_1$, $\mathcal B_2$, $\mathcal D_{WN}$ and $-\log C$ in \eqref{EUR}, let us take into account a pair of unsharp measurements $X_\eta$, $Z_\zeta$ for spin systems described by
\begin{align} 
    \hat X_{\eta_\pm}=\frac{\hat I_2\pm\eta(\sin\theta \hat\sigma_x+\cos\theta \hat\sigma_z)}{2}\nonumber \;\;\text{ and }\;\;\hat Z_{\zeta_\pm}=\frac{\hat I_2\pm\zeta \hat\sigma_z}{2}\nonumber
\end{align}
with $0\leq\eta$, $\zeta\leq1$, respectively, where $\theta$ is polar angle between directions of measurements. In this simple example, the incompatibility is determined by the angle $\theta$, and, on the other hand, the unsharpness is by $\eta$, $\zeta$. When both measurements are sharp, i.e. $\eta=\zeta=1$, the bounds $\mathcal B_1$ and $\mathcal B_2$  are reduced to the original MU bound and the bound in \eqref{MajEUR}, respectively, as depicted in Figure \ref{QDwhite}-(a). For mutually unbiased bases, i.e. $\theta=\pi/2$,  $\mathcal B_1$ is optimized to be $\log2$, whereas $\mathcal B_2$ is not. However, for $|\pi/2-\theta|>0.15$, $\mathcal B_2$ becomes stronger than $\mathcal B_1$. Comparing Figure \ref{QDwhite}-(b) to Figure \ref{QDwhite}-(a), we can find out that the measurements $Z_\zeta$ becomes unsharp, so that total device uncertainty $\mathcal D_{WN}$ increases and surpasses $-\log C$ and $\mathcal B_1$ for $|\pi/2-\theta|>0.47$ and $|\pi/2-\theta|>0.33$, respectively. Only $\mathcal B_2$ gives us stronger one than $\mathcal{D}_{WN}$. Subsequently, Figure \ref{QDwhite}-(c) shows that as unsharpness increases, $-\log C$ becomes smaller than $\mathcal D_{WN}$ over the entire range of $\theta$. With increasing the unsharpness, the gap between $-\log C$ and $\mathcal D_{WN}$ is being large as shown in Figure \ref{QDwhite}-(d). Therefore, we identify that the effect of unsharpness should be considered for nontrivial unsharp measurements, and, in particular, at sufficiently large unsharpness it becomes a dominant factor in entropic URs.}

{
\subsection*{Entropic uncertainty relation from device uncertainty }

In the above, we have seen that the unsharpness can give rise to the nontrivial lower bound $\mathcal D_{WN}$ under white noise model. Generalizing this approach, in this section, let us show that nontrivial lower bounds can be obtained for unsharp measurements by using the property of device uncertainty as following. First of all, let us consider a pair of unsharp measurements $A$ and $B$ in $\mathcal H_d$, of which the numbers of outcomes are denoted by $n$ and $m$, respectively. Then the inequalities in \eqref{single2} is directly applicable in this way, 
\begin{align} \label{DeviceEUR}
H_\rho(A)+H_\rho(B)\geq D_\rho(A)+D_\rho(B)\geq\min_\rho[D_\rho(A)+D_\rho(B)]\equiv \mathcal D,
\end{align} 
where the second inequality is obtained by diagonalizing $\left(\sum_{i=1}^{n}\sum_{k=1}^d h(a_i^k) |a_i^k\rangle\langle a_i^k| + \sum_{j=1}^{m}\sum_{l=1}^d h(b_j^l)|b_j^l\rangle\langle b_j^l|\right)$ and taking the lowest eigenvalue. Even though the bound $\mathcal D$ vanishes for sharp measurements due to the property ($D$-iii), it is considerable at sufficiently large unsharpness.

\begin{figure}[t]\centering
\includegraphics[scale =0.4]{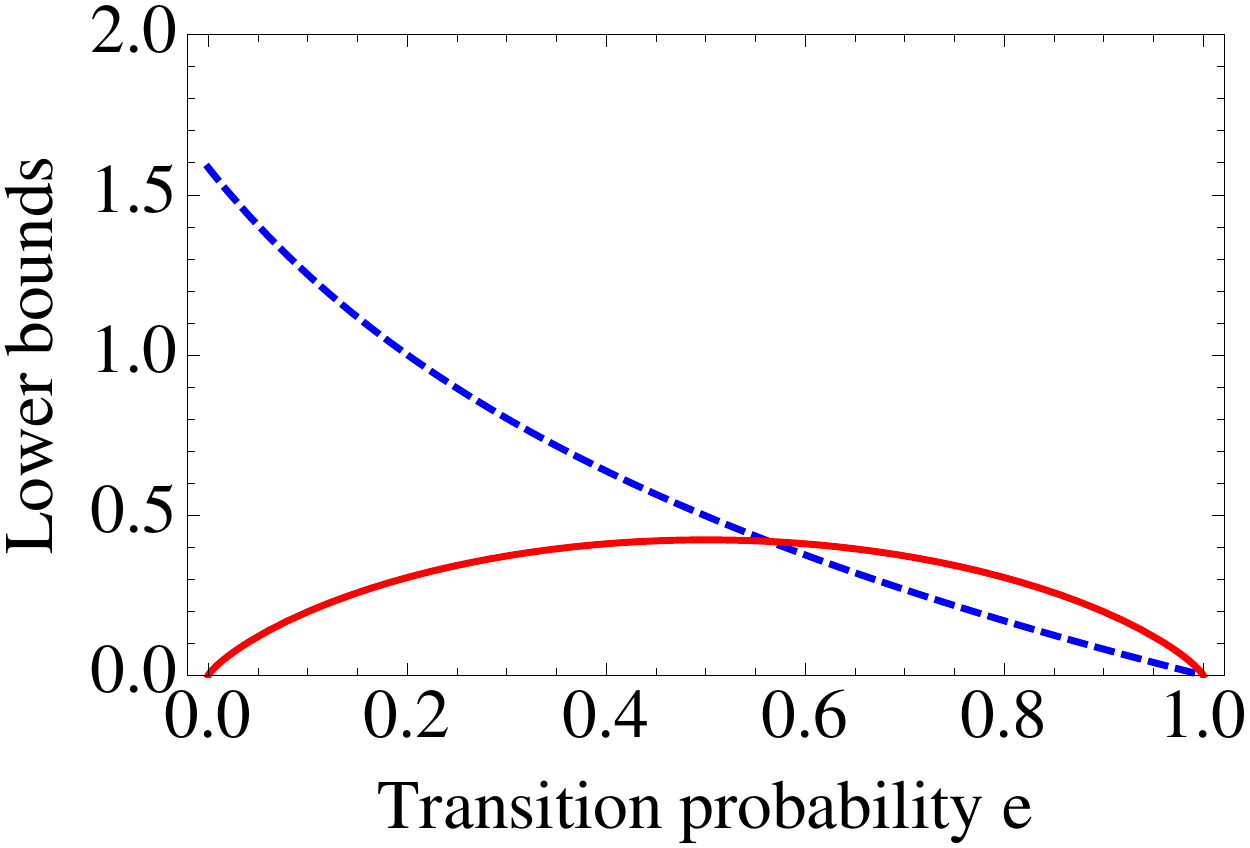} 
\caption{Graph illustrates lower bounds of entropic URs, $-\log C$ (dashed blue line) and $\mathcal D_{AD}$ (solid red line) { according to the transition probability $e$ for $X_{AD}$ and $Z_{AD}$ in 3-dimensional system. Shape of $\mathcal D_{AD}$ is given symmetrically with respect to $e$, while $-\log C$ decreases. At large $e$, $\mathcal D_{AD}$ becomes bigger than $-\log C$.}}
\label{ADcd}
\end{figure}

As an example of nontrivial noise models, let us take into account unsharp measurements $X_{AD}$ and $Z_{AD}$ constructed in $\mathcal H_3$ by adding amplitude damping noise on the measurement of mutually unbiased bases $\{|x_i\rangle\}$, $\{|z_j\rangle= 1 / \sqrt{3} \sum_{j=0}^2 e^{j2\pi i/3 } |x_j\rangle \} $ such that
\begin{align} 
    \hat X_{AD_0}=|x_0\rangle\langle x_0|+e_x |x_1\rangle\langle x_1|+e_x|x_2\rangle\langle x_2|,\nonumber\;\;
   \hat X_{AD_1}=(1-e_x )|x_1\rangle\langle x_1|,\nonumber\;\;\text{and}\;\;
    \hat X_{AD_2}=(1-e_x )|x_1\rangle\langle x_2|,\nonumber
\end{align} 
where $e_x$ is transition probability (similarly for $Z_{AD}$). 
This amplitude damping model imposes distinct noise effects with respect to states, which means that device uncertainty coming from the amplitude damping is state-dependent such that
\begin{align} 
D_\rho(X_{AD})=(p_1^{X}+p_2^{X})H_{\text{bin}}(e_x),
\end{align} 
where $p_i^X=\langle x_i|\hat \rho |x_i\rangle$ for given states $\hat \rho$. The device uncertainty, thus, vanishes for a given state $|x_0\rangle$. In the same manner, device uncertainty of $Z_{AD}$ vanishes for $|z_0\rangle$. That means each device uncertainty can vanish for specific states. Under consideration of $D_\rho(X_{AD})$ and $D_\rho(Z_{AD})$ together, however, they do not vanish simultaneously. This behavior is charactrized by means of the bound $\mathcal D$, which is given in this case by
 \begin{align} \label{DAD}
\mathcal D_{AD}=\min_\rho \big[D_\rho(X_{AD})+D_\rho(Z_{AD})\big]= \left(1-\frac{1}{\sqrt 3}\right)H_{\text{bin}}(e)
\end{align} 
with the condition $e_x=e_z=e$ for simplicity. Therefore, it is found that minimizing total device uncertainty over states can give rise to nontrivial bound, even though each device uncertainty vanishes.
Furthermore, comparing it to $-\log C$ in (\ref{EUR}) obtained as 
$-\log \left\{\frac{1}{6}\left(2+2e-e^2+3e^3+(1-e)e\sqrt{3(4+4e+3e^2)}\right)\right\}$, it is observed that $\mathcal D_{AD}$ becomes stronger than $-\log C$ when $e>0.564...$, as depicted in Fig. \ref{ADcd}. }

\section*{Discussion}
We have studied the effects of unsharpness on entropic uncertainty. As the first step, we have formulated the device uncertainty \eqref{DU} as a measure of unsharpness, and investigated its properties in line with previous works \cite{Massar2007}. The most important property among them is that the device uncertainty gives us a lower bound of the entropy by itself and also can be minimized over states as shown in \eqref{single2}.

Using the device uncertainty, we have investigated the effect of unsharp measurements in entropic URs  and observed the behavior for the lower bound to become larger by increasing the unsharpness in specific noise models such as white noise and amplitude damping. Under the white noise, device uncertainty is given as state-independent values \eqref{DC}. In that case, we have obtained two forms of lower bounds in entropic URs denoted by $\mathcal B_1$ and $\mathcal B_2$. Distinct feature they share is that they are written in the decomposed forms by discriminating the unsharpness in measuring devices. From this fact, it becomes possible to clearly observe the effect of unsharp measurement. Also, comparing these bounds to $-\log C$ in two-dimensional system, it is identified that the bounds $\mathcal B_1$ and $\mathcal B_2$ can be stronger than $-\log C$, and furthermore $-\log C$ becomes strictly weaker than the device uncertainty $\mathcal D_{WN}$ at large unsharpness as shown in Figure \ref{QDwhite}. 

In order to see the effect of unsharpness, we have derived the entropic UR in \eqref{DeviceEUR} for a pair of unsharp measurements by using the property of the device uncertainty. As a result, the lower bound $\mathcal D$ originating from the unsharpness has been obtained and compared to $-\log C$  in a specific example to show its validity as a nontrivial bound. As one of noise models, we take into account the amplitude damping model in three-dimensional system. In this case, each device uncertainty of the unsharp measurement is state-dependent and vanishes for specific states. A notable point, nevertheless, is that the nonzero bound $\mathcal D_{AD}$ in \eqref{DAD} has been obtained from the relation in \eqref{DeviceEUR}. Furthermore, it is observed that $-\log C$ in (\ref{EUR}) becomes smaller than $\mathcal D_{AD}$ at sufficiently large unsharpness as shown in Figure \ref{ADcd}. Conclusively, the results so far consistently show that the effect of the unsharpness in the entropic UR is considerably large, and thus should be taken into account to develop the entropic UR for unsharp measurements. More generalization of the uncertainty relation can be studied when we consider more than two measurements for a single system. We leave them to a future investigation.

\section*{Methods}

\subsection*{The proof of the second relation in (\ref{single2})}

{\bf The second relation in (\ref{single2})}
The device uncertainty is larger than the lower bound of (\ref{single}),
\begin{equation}
D_\rho(A)\geq -\log\left[\max_i\|\hat A_i\|\right]
\end{equation}
for all states $\rho$.

{ Proof}. The device uncertainty can be written as for any POVM 
\begin{align*}
D_\rho(A)=-\sum_i^n\left(\sum_{k=1}^d \langle a_i^k|\hat \rho| a_i^k\rangle a_i^k \log a_i^k\right),
\end{align*}
where $d$ is the dimension of Hilbert space, and $n$ is the number of elements of POVM. Due to $-\log x$ is decreasing function for positive $x$, we can find out
\begin{align}
D_\rho(A)&\geq-\sum_i^n\left(\sum_{k=1}^d \langle a_i^k|\hat \rho| a_i^k\rangle a^i_k \log (\max_{i,k} a^i_k)\right)\nonumber=-\log (\max_{i,k} a_i^k)\sum_i^n\sum_{k=1}^d \langle a_i^k|\hat \rho| a_i^k\rangle a_i^k \nonumber\\
&=-\log (\max_{i,k} a_i^k)\sum_i^n \text{Tr}[\hat \rho \hat A_i]\nonumber=-\log (\max_{i,k} a_i^k)=-\log\left[\max_i\|\hat A_i\|\right]\nonumber.
\end{align}
The last line comes from the completeness relation satisfied by POVM.

\subsection*{Proofs of the conditions (D-i)-(D-iv)}
We will show the proof of the conditions (D-i)-(D-v) for the general device uncertainty (\ref{DU}) in consecutive order.

{\bf Condition (D-i)} $0\leq  D_\rho(A)\leq H_\rho(A)$.

{Proof.} Entropy of probability distribution $\{p_i\}$ of outcomes of measurement $A$ described by POVM $\{\hat A_i\}$ is written by
\begin{align}
H_\rho(A)=-\sum_i p_i \log p_i,
\end{align}
where the probability of $i$th outcome is given by
\begin{align}
p_i=\text{Tr}[\hat \rho \hat A_i ]=\sum_{k=1}^d a_i^k \langle a_i^k|\hat \rho| a_i^k\rangle,
\end{align}
with decomposed form of positive operator $\hat A_i=\sum_{k=1}^d a_i^k |a_i^k\rangle\langle a_i^k|$. Since eigenvectors $|a_i^k\rangle$ of $\hat A_i$ satisfy the completeness relation, $p_i$ is written as the convex combination of eigenvalues $a_i^k$'s. Therefore, using the concavity of $-x\log x$ \cite{Cover1991},
\begin{align}
H_\rho(A)=\sum_i^n- p_i \log p_i \geq -\sum_i^n\sum_{k=1}^d  \langle a_i^k|\hat \rho| a_i^k\rangle a_i^k \log a_i^k=D_\rho(A).
\end{align}
In addition, according to the restriction of POVM, i.e. completeness and positivity, the device uncertainty should be positive. As a result, it is proved
\begin{align}
0\leq D_\rho(A)\leq H_\rho(A).
\end{align}
QED.

{\bf Condition(D-ii)} $D_\rho(A)=H_\rho(A)$ for all states if and only if all positive operators of $A$ is in the form of $\hat A_i=\lambda_i \hat I_d$ with $0\leq\lambda_i\leq1$ satisfying $\sum_{i=1}^n \lambda_i=1$.

{Proof.} The left direction of proof is trivial. Thus let us prove the right direction. If $D_\rho(A)=H_\rho(A)$ for all states, then we can obtain the following equation
\begin{align}
H_\rho(A)-D_\rho(A)\label{D-ii}
=\sum_{i=1}^n\left(-\sum_{k=1}^d   \langle a_i^k|\hat \rho| a_i^k\rangle a_i^k \log \frac{\left(\sum_{l=1}^d   \langle a_i^k|\hat \rho| a_i^k\rangle a_i^l\right)}{a_i^k}\right)=0\nonumber
\end{align}
for all sates. Then each $i$th term in the bracket is always positive due to the concavity of the function, $-x\log x$. Thus all $i$th term should be zero. Then we can choose the input state to be a superposition of two eigenstates of $\hat A_i$, since it should be satisfied for all states. For example, in a case we take it as $\hat \rho=|\psi\rangle\langle\psi|$, where $|\psi\rangle=(|a_i^1\rangle+|a_i^2\rangle)/\sqrt{2}$, $i$the term becomes simple form and it should be zero such as
\begin{align}
-\frac{1}{2} a_i^1 \log\left(\frac{a_i^1+a_i^2}{2a_i^1}\right)-\frac{1}{2} a_i^2 \log\left(\frac{a_i^1+a_i^2}{2a_i^2}\right)=0.
\end{align}
As taking partial derivartive of $a_i^1$, we can obtain the minimum value of the left hand side at $a_i^1=a_i^2$, and its value is given by 0. Otherwise, it is always strictly positive. Thus for (\ref{D-ii}) to vanish, $a_i^1$ and $a_i^2$ should be same. Then as taking the input state to be superposition of all combination of eigenstates, it can be shown that all eigenvalues $a_i^k$'s must be same, which means $\hat A_i$ is in the form of $\lambda_i \hat I_d$.
QED

{\bf Condition(D-iii)} $D_\rho(A)=0$ for all states if and only if $A$ is PVM.

{Proof.} The observable $A$ is PVM if and only if eigenvalues of all operators, $\{\hat A_i=\sum_{k=1}^d a_i^k |a_i^k\rangle\}$, should be given by 0 or 1, i.e. $a_i^k\in\{0,1\}$. And $D_\rho(A)=0$ for all states is also equivalent with the condition all $a_i^k\in\{0,1\}$, since the device uncertainty vanish for all states means $-a_i^k\log a_i^k=0$ for all $i$, $k$.
QED

{\bf Condition(D-iv)} A convex combination of two POVMs cannot be sharper than these two POVMs themselves.

{ Proof.} A convex combination of two POVMs $A$ and $B$, acting on same quantum system described by $\hat \rho$ with probabilities $p$ and $1-p$ satisfying $0\leq p\leq 1$ respectively, is denoted by
\begin{align}
\{p\hat A_i, (1-p)\hat B_j\}.
\end{align}
The decomposed form of the convex combination of $A$ and $B$ is written by
\begin{align}
\{\sum_{k=1}^d p\:a_i^k|a_i^k\rangle\langle a_i^k|, \sum_{l=1}^d (1-p)b_j^l|b_j^l\rangle\langle b_j^l|\}.
\end{align}
Thus eigenvalues of each element of $A$ and $B$ are multiplied by $p$ and $(1-p)$, respectively. That means the device uncertainty of combined POVM is given by
\begin{align}
&D_\rho(pA+(1-p)B)\\
&=-\sum_{i=1}^n \sum_{k=1}^d \langle a_i^k|\hat \rho| a_i^k\rangle (pa_i^k)\log (pa_i^k)\nonumber-\sum_{j=1}^m \sum_{l=1}^d \langle b_j^l|\hat \rho| b_j^l\rangle ((1-p)b_j^l)\log ((1-p)b_j^l)\nonumber\\
&=-p\sum_{i=1}^n \sum_{k=1}^d \langle a_i^k|\hat \rho| a_i^k\rangle  a_i^k\log a_i^k -p\log p\nonumber-(1-p)\sum_{j=1}^m \sum_{l=1}^d \langle b_j^l|\hat \rho| b_j^l\rangle b_j^l\log b_j^l-(1-p)\log(1-p)\nonumber\\
&=pD_\rho(A)+(1-p)D_\rho(B)+H_{\text{bin}}(p)
\end{align}
As a result, the convex combination of two POVMs increases by $H_{\text{bin}}(p)$.QED

\subsection*{Proof of (\ref{Qwn})}\label{ProofQmub}
Let us consider sharp observables $A$ and $B$ in $\mathcal H_d$ associated with the orthonormal bases $\{|a_i\rangle\}$ and $\{|b_j\rangle\}$, respectively. And we assume that white noise acts on $A$ such as (\ref{WN}) and similarly for $B$. Consequently, $A$ and $B$ become unsharp observables $A_{\alpha}$ and $B_\beta$, respectively.

Before proving (\ref{Qwn}), we define entropic URs for sharp measurements in mixed states $\hat\rho$ as
\begin{align}\label{EURberta}
H_\rho(A)+H_\rho(B)\geq -\log c_{ab}^2+S(\rho)
\end{align}
with the von Neumann entropy $S(\rho)=-\text{Tr}(\rho\log\rho)$,
which is the reduced form of the entropic UR derived for bipartite system \cite{Berta2010} . According to dual map, we have a relation such that 
\begin{align}
\langle a_i|\hat\rho_\alpha|a_i\rangle=\langle \psi|(\hat A_\alpha)_i|\psi\rangle
\end{align}
with definitions $\hat\rho_\alpha=\alpha |\psi\rangle\langle\psi| +(1-\alpha)\hat I_d/d$ and $\hat (A_\alpha)_i=\alpha |a_i\rangle\langle a_i| +(1-\alpha)\hat I_d/d$.
In this case, device uncertainty of $A_\alpha$ have the same value with von Neumann entropy of $\rho_\alpha$, i.e. $S(\rho_\alpha)=D(A_\alpha)$. Using these relations, the relation (\ref{EURberta}) is rewritten in the from of 
\begin{align}
H_\psi(A_\alpha)+H_\psi(B_\alpha)\geq -\log c_{ab}^2+D(A_\alpha)
\end{align}
for the case of white noise.
With the fact that $H(A_\alpha)\geq H(A_\beta)$ if $\alpha\leq\beta$, we have following result 
\begin{align}
H_\psi(A_\alpha)+H_\psi(B_\beta)\geq -\log c_{ab}^2+\min[D(A_\alpha),D(B_\beta)],
\end{align}
where the right hand side coincides with $\mathcal B_1$. QED


\section*{Acknowledgements}
This work was done with support of ICT R$\&$D program of MSIP/IITP (No.2014-044-014- 002), the R\&D Convergence Program of NST of Republic of Korea (Grant No. CAP-15-08-KRISS). and National Research Foundation (NRF) grant (No.NRF- 2013R1A1A2010537).

\section*{Author contributions statement}
W. Son suggest the major formalism and K. Baek performed detailed calculation. K. Baek and W. Son contribute to the manuscript preparation more or less equally.

\section*{Additional information}

{\bf Competing financial interests}: The authors declare no competing financial interests.

\noindent
{\bf How to cite this article}: Baek, K. and Son, W. Effects of generalized measurement on entropic uncertainty relation. Sci. Rep. *, *; doi: */ srep* (2016).

\end{document}